\documentclass[fleqn,12pt]{article}
 \usepackage{epsfig}
\usepackage{amssymb}
\def\cal{\fam2 }
\newcommand{\ba}{\begin{eqnarray}}

\newcommand{\ea}{\end{eqnarray}}
\newcommand{\be}{\begin{equation}}
\newcommand{\ee}{\end{equation}}

\newcommand{\eq}[1]{Eq.\,(\ref{#1})}

\newcommand{\pbar}{\bar p}

\newcommand{\delchi}{\Delta \chi^2_i}

\newcommand{\delchimax}{{\delchi}_{\rm max}}
\newcommand{\x}{(\nu/m)}
\newcommand{\y}{(\nu_0/m)}

\def\bea{\begin{eqnarray}} 
\def\eea{\end{eqnarray}} 
\def\rd{{\mathrm d}} 
 
\def\intd4x{\int{\rd}^4x}

\def\m32{{m_{3/2}}}

\begin{document}
\renewcommand\thepage{\ }
\begin{titlepage} 
%
\newcommand\mydate{\today} 
\newlength{\nulogo} 
\settowidth{\nulogo}{\small\sf{NUHEP Report XXXX}}
\title{
\vspace{-.8in} 
\hfill
         {\small\sf \mydate}
\vspace{0.5in} \\
{
Implications from analyticity constraints  used in a Landshoff-Donnachie fit
}}

\author{
M.~M.~Block\\
{\small\em Department of Physics and Astronomy,} \vspace{-5pt} \\ 
{\small\em Northwestern University, Evanston, IL 60208}\\
\vspace{-5pt}
\  \\
F.~Halzen
\vspace{-5pt} \\ 
{\small\em Department of Physics,} 
\vspace{-5pt} \\ 
{\small\em University of
Wisconsin, Madison, WI 53706} \\
\vspace{-5pt}\\
%
\vspace{-5pt}\\
%
}    
\vspace{.5in}
\vfill
\date {}
\maketitle
\begin{abstract}

Landshoff and Donnachie[hep/ph 0509240, (2005)] parametrize the energy behavior of $pp$ and $p\bar p$ scattering cross sections with five parameters, using:
\ba
\sigma^+&=&56.08 s^{-0.4525}+21.70s^{0.0808}\quad {\rm for \ }pp,\\
\sigma^-&=&98.39 s^{-0.4525}+21.70s^{0.0808}\quad {\rm for \ }p \bar p.
\ea 
Using the 4 analyticity constraints of Block and Halzen[M. M. Block and F. Halzen, Phys. Rev. D {\bf 72}, 036006 (2005)], we simultaneously fit the Landshoff-Donnachie form to the same ``sieved'' set of $pp$ and $p\bar p$ cross section and $\rho$ data that Block and Halzen  used for a very good fit to a $\ln^2 s$ parametrization. We show that the satisfaction of the analyticity constraints will require complicated modifications of the Landshoff-Donnachie parametrization for lower energies, greatly altering its inherent appeal of simplicity and universality.
\end{abstract}
\end{titlepage} 
\renewcommand{\thepage}{\arabic{page}} 
Landshoff and Donnachie\cite{landshoff0,landshoff1,landshoff2} parameterize the scattering cross section with five parameters, using:
\ba
\sigma^+&=&56.08 s^{-0.4525}+21.70s^{0.0808}\quad {\rm for \ }pp,\label{pp}\\
\sigma^-&=&98.39 s^{-0.4525}+21.70s^{0.0808}\quad {\rm for \ }p \bar p,\label{ppbar}
\ea
where $s$ is in GeV$^2$. 
Using the high energy limit $s\rightarrow 2 m \nu$, where $\nu$ is the laboratory energy and $m$ is the proton mass, we rewrite these equations, as well as $\rho$, the ratio of the real part to the imaginary part of the forward scattering amplitude,  more generally as
\ba
\sigma^\pm&=&A\left(\frac{\nu}{m}\right)^{\alpha-1}    +B\left(\frac{\nu}{m}\right)^{\beta-1}\pm D \left(\frac{\nu}{m}\right)^{\alpha-1}\label{sigma+-},\\
\frac{d\sigma^\pm\ \ }{d\x}&=&A(\alpha -1)\left(\frac{\nu}{m}\right)^{\alpha-2}+B(\beta -1)\left(\frac{\nu}{m}\right)^{\beta-2}\pm D(\alpha -1)\left(\frac{\nu}{m}\right)^{\alpha-2}\label{deriv+-},\\
\rho^\pm&=&{1\over\sigma^\pm}\left\{-A \cot\left({\pi\alpha\over 2}\right)\left({\nu\over m}\right)^{\alpha -1} - B \cot\left({\pi\beta\over 2}\right)\left({\nu\over m}\right)^{\beta-1}+\frac{4\pi}{\nu}f_+(0)\right.\nonumber\\
&&\qquad\qquad\qquad\qquad\qquad\qquad\left.
\pm \ D\tan\left({\pi\alpha\over 2}\right)\left({\nu\over m}\right)^{\alpha -1} \right\}\label{rho+-},\ea
where $m$ is the proton mass,  the upper sign is for pp and the lower for $p\bar p$ scattering. We have used the analyticity properties\cite{bc} of real analytic amplitudes to write  $\rho^\pm$ in \eq{rho+-}. The  6 real parameters which are needed are:  3 Regge coefficients, $A, B$ and $D $ in mb, 2 Regge powers, $\alpha$ and $\beta$, which are  dimensionless and $f_+(0)$.  The  real constant $f_+(0)$ introduced in \eq{rho+-} is the subtraction constant at $\nu=0$ needed for a singly-subtracted dispersion relation\cite{{bc},{gilman}}.

Using \eq{pp} and \eq{ppbar}, along with \eq{sigma+-},  we find  $\alpha=0.5475$ and $\beta=1.0808$, with $A=59.8$ mb, $B=22.71$ and $D=-16.38$ mb, where the energy variable is now $\nu/m$, instead of $s$.

Let us now consider a transition energy $\nu_0$, defined  as an energy slightly higher than the energy where the resonances average out, i.e., an energy where the cross sections already have a smooth behavior (a useful choice for $p \bar p $ and $pp$ reactions is $\nu_0=7.59$ GeV, corresponding to the c.m. (center-of-mass) energy $\sqrt s_0= 4 $ GeV).  At the transition energy $\nu_0$, it is convenient to define the 4 analyticity conditions\cite{{block},{bh}}
\begin{eqnarray}
\sigma_{\rm av}&=&\frac{\sigma^{+}(\nu_0)+\sigma^-(\nu_0)}{2}\qquad\quad
=\quad A\y^{\alpha -1}+B\y^{\beta -1},\label{sigav}\\
\Delta\sigma&=&\frac{\sigma^{+}(\nu_0)-\sigma^-(\nu_0)}{2}\qquad\quad
=\quad D\y^{\alpha -1},\\
m_{\rm av}&=&\frac{1}{2}\left[\frac{d\sigma^{+}}{d\left(\frac{\nu}{m}\right)}+\frac{d\sigma^{-}}{d\left(\frac{\nu}{m}\right)}\right]_{\nu =\nu_0}
= \quad A(\alpha -1)\y^{\alpha - 2}+B(\beta -1)\y^{\beta - 2},\\
\Delta m&=&\frac{1}{2}\left[\frac{d\sigma^{+}}{d\left(\frac{\nu}{m}\right)}-\frac{d\sigma^{-}}{d\left(\frac{\nu}{m}\right)}\right]_{\nu =\nu_0}
=\quad D(\alpha -1)\y^{\alpha - 2}.\label{delm}
\end{eqnarray}
Using these definitions of the experimental quantities $\sigma_{\rm av}$, $\Delta\sigma$, $m_{\rm av}$ and $\Delta m$, we now write the four analyticity constraints at energy $\nu_0$, using  \eq{sigma+-} and \eq{deriv+-}---see references \cite{block} and\cite{bh}---in terms of the one free parameter $A$, 
\begin{eqnarray}
\alpha&=&1+\frac{\Delta m}{\Delta \sigma}\y,\label{derivodd}\\
D&=&\Delta \sigma\y^{1-\alpha}\label{interceptodd},\\
\beta(A)&=&1+\frac{m_{\rm av}\y-A(\alpha -1)\y^{\alpha -1}}{\sigma_{\rm av}-A\y^{\alpha -1}},\label{betaofA}\\
B(A)&=&\sigma_{\rm av}\y^{1-\beta}-A\y^{\alpha-\beta}.\label{BofA}
\end{eqnarray}
These  analyticity consistency conditions\cite{block} utilize the two experimental cross sections and their first derivatives at the transition energy $\nu_0$, where we join on to the asymptotic fit. We have chosen  $\nu_0$ as the (low) energy just after which resonance behavior finishes.  At $\sqrt s_0 =4$ GeV (corresponding to $\nu_0=7.59$ GeV), Block and Halzen\cite{bh} found  that
\ba
\sigma^+(\nu_0)&=&40.18 \quad\quad{\rm mb,}
\qquad \sigma^-(\nu_ 0)\quad=\quad\quad 56.99  \quad\quad{\rm mb,}\label{sigs}\\
\left.\frac{d\sigma^+\ \ }{d\x}\right|_{\nu =\nu_0}&=&-0.2305 \quad{\rm mb,}\quad
\left.\frac{d\sigma^-\ \ }{d\x}\right|_{\nu=\nu_0}=-1.4456\quad{\rm mb,}
\ea
using a local fit in the neighborhood of $\nu_0$. 

These values yield the 4 constraints required by analyticity, i.e., 
\begin{eqnarray}
\sigma_{\rm av}(\nu_0)&=&48.59 \quad\quad\,{\rm mb},\qquad
\Delta\sigma(\nu_0)=\quad -8.405 \quad{\rm\  mb},\label{delsig&sigav}\\
m_{\rm av}(\nu_0)&=&-0.8381\quad{\rm mb},\qquad
\Delta m(\nu_0)=\quad 1.215\,\, \quad{\rm \ mb}\label{Delm&mav}.
\end{eqnarray}

Using the numerical values in \eq{delsig&sigav} and  \eq{Delm&mav} for the odd amplitude, along with \eq{derivodd} and \eq{interceptodd}, we note  that the odd amplitude is {\rm completely} specified. This is true even {\em before} we make a fit to the high energy data. The two odd analyticity conditions constrain the odd parameters to be
\ba
D&=&-28.56\quad{\rm mb},\quad\quad\,\quad D_{\rm LD}\quad=\quad-16.38\quad{\rm mb}, \label{delta}\\
\alpha&=&0.4150,\quad
\qquad\qquad\quad\alpha_{\rm LD}\quad=\quad0.545\label{alpha}.
\ea
where we have contrasted  these value with the values found by Landshoff and Donnachie, $D_{\rm LD}$ and $\alpha_{\rm LD}$, which are clearly incompatible.

Next, we use the two constraint equations, \eq{betaofA} and \eq{BofA}, along with \eq{delta} and \eq{alpha}, together with the even amplitude portions of \eq{delsig&sigav} and \eq{Delm&mav}, to simultaneously fit a ``sieved'' data set\cite{sieve,bh} of high energy cross sections and $\rho$-values for $pp$ and $p\bar p$ with energies above $\sqrt s =6 $ GeV, derived from the archives of the Particle Data Group\cite{pdg}. The ``sieve'' algorithm which was used to find this data set is fully described in ref. \cite{sieve}.  

This same data set has already been successfully used to make an excellent $\ln^2$ fit\cite{bh} of the type  used in \eq{lnsqs}, using the {\em same} analyticity constraints\cite{bh} as we use here. It should be further noted that a $\ln s$ fit, i.e., setting the coefficient $c_2=0$ in \eq{lnsqs}, again using the identical analyticity constraints of \eq{delsig&sigav} and \eq{Delm&mav} as well as the same ``sieved'' data set, was conclusively ruled out\cite{bh}. 

After employing the 4 constraints, the number of fit parameters has been reduced from 6 to 2, i.e., the two free parameters $A$ and $f_+(0)$. It should be noted that the subtraction constant $f_+(0)$ only enters into $\rho^\pm$-values and {\em not} into  cross section determinations $\sigma^\pm$. In essence, the cross section fit is a one-parameter fit, $A$.

\begin{figure}[] 
\begin{center}
\mbox{\epsfig{file=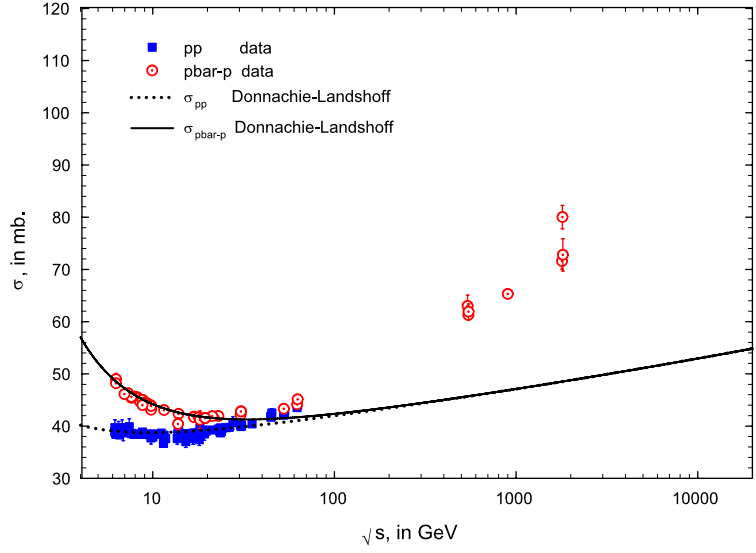,width=4in%
,bbllx=0pt,bblly=0pt,bburx=411pt,bbury=325pt,clip=%
}}
\end{center}
\caption[]{ \footnotesize
The fitted total cross sections $\sigma_{p p}$ and $\sigma_{\pbar p}$ in mb, {\em vs.} $\sqrt s$, in GeV, using the 4 constraints of Equations (\ref{delsig&sigav}) and (\ref{Delm&mav}). The circles are the sieved data\cite{bh,sieve}  for $\pbar p$ scattering and the squares are the sieved data\cite{bh,sieve}  for $p p$ scattering for c.m. energies  $\sqrt s\ge 6$ GeV. The solid curve ($\pbar p$) and  the dotted curve ($\pbar p$) are the  $\chi^2$ fits from Table \ref{table:Landshoff}.
}
\label{fig:sigmapp}
\end{figure}

\begin{figure}[] 
\begin{center}
\mbox{\epsfig{file=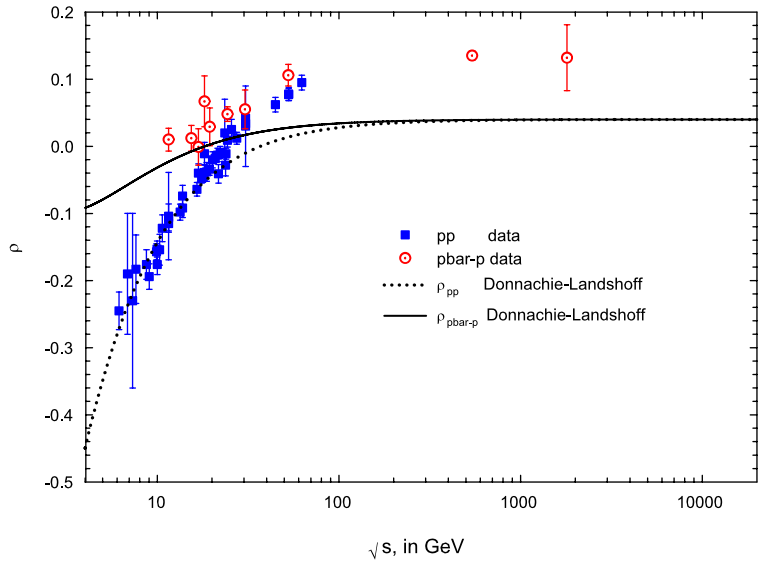,width=4in%
,bbllx=0pt,bblly=0pt,bburx=411pt,bbury=325pt,clip=%
}}
\end{center}
\caption[]{ \footnotesize
The fitted $\rho$-values, $\rho_{p p}$ and $\rho_{\pbar p}$, {\em vs.} $\sqrt s$, in GeV, using the 4 constraints of Equations (\ref{delsig&sigav}) and (\ref{Delm&mav}).  The circles are the sieved data\cite{bh,sieve}  for $\pbar p$ scattering and the squares are the sieved data\cite{bh,sieve}  for $p p$ scattering for c.m. energies $\sqrt s\ge 6$ GeV. The solid curve ($\pbar p$) and  the dotted curve ($\pbar p$) are the  $\chi^2$ fits from Table \ref{table:Landshoff}.
  }
\label{fig:rhopp}
\end{figure}

The results of the fit are given  in Table \ref{table:Landshoff}. 
When the cross sections using the parameters of Table \ref{table:Landshoff} are rewritten in terms of $s$, for direct comparison with the Landshoff-Donnachie cross sections of \eq{pp} and \eq{ppbar}, we find that our analyticity-constrained cross sections are: 
\ba
\sigma^+&=&23.97 s^{-0.5850}+33.02s^{0.0255}\quad {\rm for \ }pp,\label{pp1}\\
\sigma^-&=&109.1 s^{-0.4525}+33.02s^{0.0255}\quad {\rm for \ }p \bar p\label{pbarp1}.
\ea
Clearly, our \eq{pp1} and \eq{pbarp1} are in sharp disagreement with the Landshoff-Donnachie cross sections given in \eq{pp} and \eq{ppbar}. This is graphically seen in Figures \ref{fig:sigmapp} and \ref{fig:rhopp}, which plot $\sigma$ and $\rho$, respectively,  against the c.m.  energy $\sqrt s$, where we see that the fits are very much below {\em all} of the high energy points. 
The renormalized\cite{sieve} $\chi^2$ per degree of freedom is 21.45, for 185 degrees of freedom, yielding the incredibly large value of 3576.24 for the total $\chi^2$. 

Thus, there is essentially zero probability that a fit of the Landshoff-Donnachie type---of the form given in \eq{pp} and \eq{ppbar}---is a good representation of the high energy data ($\sqrt s\ge 6$ GeV). Certainly, at the very high energy end, their  functional form  violates unitarity. We now see that it does not have the proper shape to satisfy analyticity  at the lower energy end.  Clearly, the form  requires substantial ad hoc modifications to join on to the low energy constraints. Thus its primary virtue---its simplicity of form---requires serious modification. 
We bring to the reader's attention that a $\ln^2s$ fit of the form
\be
\sigma^\pm(\nu)=c_0 +c_1\ln(\nu/m)+c_2\ln^2(\nu/m)+\beta_{\cal P'}(\nu/m)^{-.5}\pm \delta (\nu/m)^{\alpha -1},\label{lnsqs}
\ee 
was carried out on the {\em same} ``sieved'' sample of $\sigma^\pm$ and $\rho^\pm$  in ref. \cite{{bh},{bh2}}, using the {\em same} 4 analyticity constraints, where it gave a renormalized $\chi^2$ per degree of freedom of 1.095 for 184 degrees of freedom, an excellent fit. Further, this $\ln^2$ type fit was shown to be independent of the choice of transition energy $s_0$\cite{{bh},{bh2}}, for $4\le s_0\le 6$ GeV.

In conclusion,  a functional form of the type
\ba
\sigma(pp)&=&A's^{\alpha-1}+B's^{\beta-1},\\
\sigma(p\bar p)&=&C's^{\alpha-1}+B's^{\beta-1},
\ea 
with $\beta \sim 1.1$, although conceptually very simple,
can not be used for fitting high energy scattering for energies $\sqrt s>6$ GeV, since it can not satisfy the 4 analyticity requirements of Equations (\ref{delsig&sigav}) and (\ref{Delm&mav}). In addition, the term $s^{\beta -1}$ violates unitarity at the highest energies. Thus, this simple type of parametrization---which was widely used because of its inherent simplicity---is effectively excluded, since it now requires {\em substantial} modification to its low energy behavior. We suspect that the earlier success of the Landshoff-Donnachie model reflected its validity in only a very limited energy region. In contrast,  the $\ln^2s$ parametrization of Block and Halzen\cite{bh} gives an excellent fit, satisfying unitarity in a natural way, as well as satisfying  the 4 analyticity constraints.

The work of FH is supported  in part by the U.S.~Department of Energy under Grant No.~DE-FG02-95ER40896 and in part by the University of Wisconsin Research Committee with funds granted by the Wisconsin Alumni Research Foundation. One of us (MMB) would like to thank the Aspen Center of Physics for its hospitality during the writing of this paper.  

\begin{table}[h]                   
%
\begin{center}
\def\arraystretch{1.2}            
\vspace{.5in}
     \caption{\protect\small The fitted results for a 2-parameter fit of the Landshoff-Donnachie type, ($\sigma^\pm=A(\nu/m)^{\alpha-1} +B(\nu/m)^{\beta-1}\pm D(\nu/m)^{\alpha-1}$, where $\nu$ is the laboratory projectile  energy and $m$ is the proton mass), simultaneously to both the total cross sections and $\rho$-values for $pp$ and $p \bar p$ scattering, for c.m. energies $\sqrt s \ge 6$ GeV, using the sieved data set of Ref. \cite{{bh},{bh2}}.  The renormalized $\chi^2_{\rm min}$ per degree of freedom,  taking into account the effects of the $\delchimax=6$ cut\cite{sieve}, is given in the row  labeled ${\cal R}\times\chi^2_{\rm min}$/d.f. The errors in the fitted parameters have been multiplied by the appropriate $r_{\chi2}$\cite{sieve}.    \label{table:Landshoff}}. 

\begin{tabular}[b]{|l||c||}
\hline
{Parameters}&Even Amplitude\\
      \hline
      $A$\ \ \   (mb)&$44.65\pm 0.0031$ \\ 
      $B$\ \ \   (mb)&$33.60$ \\ 
	$\beta$&$1.0255$\\
	$f_+(0)$ (mb GeV)&$2.51\pm 0.57$\\
      \hline
{}&Odd Amplitude\\
	\hline
      $D$\ \ \   (mb)&$-28.56$\\
      $\alpha$&$0.415$ \\ 
	\cline{1-2}
     	\hline
	\hline
	$\chi^2_{\rm min}$&3576.2\\
	${\cal R}\times\chi^2_{\rm min}$&3968.1\\ 
	d.f.&185\\
\hline
	${\cal R}\times\chi^2_{\rm min}$/d.f.&21.45\\
\hline
\end{tabular}
\end{center}
\end{table}
\def\arraystretch{1}  
\newpage
%
\end{document}